\documentclass[aps,prb,twocolumn,groupedaddress,showpacs]{revtex4}
\usepackage{graphicx}
\usepackage{amssymb}
\usepackage{amsmath}
\usepackage{dcolumn}
\usepackage{color}

\begin{document}

\title{Electronic and phononic properties of cinnabar: ab initio calculations and some experimental results}

\author{M. Cardona}
\author{R. K. Kremer}
\email[Corresponding author:~E-mail~]{r.kremer@fkf.mpg.de}
\author{R. Lauck}
\author{G. Siegle}
\affiliation{Max-Planck-Institut f{\"u}r Festk{\"o}rperforschung,
Heisenbergstrasse 1, D-70569 Stuttgart, Germany}

\author{A. Mu\~{n}oz}
\affiliation{MALTA Consolider Team, Departamento de F\'{\i}sica Fundamental II, and Instituto de Materiales y Nanotecnolog\'{\i}a, Universidad de La Laguna, La Laguna 38205, Tenerife, Spain}

\author{A.H. Romero}
\affiliation{CINVESTAV, Departamento de Materiales, Unidad
Quer$\acute{e}$taro, Quer$\acute{e}$taro, 76230, Mexico}

\author{M. Schmidt}
\affiliation{Max-Planck-Institut f{\"u}r Chemische Physik fester Stoffe, N\"othnitzerstr. 40, D-01187 Dresden, Germany}

\date{\today}

\begin{abstract}
We report \textit{ab initio} calculations of the electronic band structure, the corresponding optical spectra, and the phonon dispersion relations of trigonal $\alpha$-HgS (cinnabar). The calculated dielectric functions are compared with unpublished optical measurements by Zallen and coworkers. The phonon dispersion relations are used to calculate the temperature and isotopic mass dependence of the specific heat which has been compared with experimental data obtained on samples with the natural isotope abundances of the elements Hg and S (natural minerals and vapor phase grown samples) and on samples prepared from isotope enriched elements by vapor phase transport. Comparison of the calculated vibrational frequencies with Raman and ir data is also presented.  Contrary to  the case of cubic $\beta$-HgS (metacinnabar), the spin-orbit splitting of the top valence bands at the $\Gamma$-point of the Brillouin zone ($\Delta_0$ ) is positive, because of a smaller admixture of 5$d$ core electrons of Hg. Calculations of the lattice parameters, and the pressure dependence of $\Delta_0$ and the corresponding direct gap $E_0 \sim$ 2eV  are also presented. The lowest absorption edge is confirmed to be indirect.
\end{abstract}

 \maketitle


\email{M.Cardona@fkf.mpg.de}

\section{Introduction}\label{SecIntro}

Cinnabar ($\alpha$-HgS) is the main ore for the production of mercury and, in powdered form, constitutes the red pigment vermillion which was already used in pre-Columbian Peru as early as 500 BC (Chavin Empire).  Large scale mining of cinnabar is known to have taken place after the conquest of the Inca Empire (1532 AD) in connection with the extraction of silver from low grade ores. It probably led to the first pre-industrial source of Hg environmental pollution.~\cite{Cooke2009}
Evidence for the use of cinnabar as a pigment is also found in Mesoamerica, dating back to the Olmec culture ($\sim$800 BC), where it was utilized in ceremonial burials and for coloring beautiful ceramic figurines.~\cite{Poole2007}  It was mined in Quer\'{e}taro, where nowadays the Institute of one of the present authors (AHR) is located.

In the Eastern World, China, today the main producer of mercury, was early using cinnabar as a pigment. The best known use is found in the lacquerware of the Song Dynasty (1000 AD).~\cite{Crick2009}  It is applied, up to now, in traditional Chinese medicine (as Zhu Sha) to treat a variety of ailments including colds, insomnia, restlessness and, less dangerously externally, skin disorders. The interested reader will find in the web abundant therapeutic as well as ordering information.

The extraction of mercury from cinnabar is documented in Teophrastus of Eresus' "Book on Stones" ($\sim$315 BC)~\cite{Takacs2000}: "Native cinnabar was rubbed with vinegar in a copper mortar  with a copper pestle" , thus describing what is probably the first mechano-chemical reaction. Pliny the Elder (23-79 AD), in his natural history~\cite{Plinius1961}, describes not only the mechano-chemical but also the distillation method which seems to have originated from Dioscorides (40-90 AD).

Because of the existence of large cinnabar deposits, vermillion was also used to illuminate ancient manuscripts in Spain and Italy. The largest world deposits are probably still those of Almad\'{e}n in Spain but the mines were closed in 2002 because of the drop in the price of mercury. The largest producers are now China and Kyrgyztan.

Beautiful red single crystals of $\alpha$-HgS can be pried out of mineral samples. It is also possible to grow them from the elements using conventional vapor phase techniques. A method to grow thin films on substrates for possible optoelectronic applications was disclosed in 1974.~\cite{RockwellPat}

Although cinnabar is the stable form of HgS under NTP, a zincblende-type modification, metacinnabar ($\beta$-HgS), is also found in nature and can be grown in the laboratory by adding traces of iron ($\sim$1\%) to the elements used in the growth. The electronic and vibronic properties of metacinnabar have been discussed recently.~\cite{Cardona2009}

The present article follows the structure of Ref. \onlinecite{Cardona2009}, with emphasis on the differences between the crystal structure of cinnabar and that of metacinnabar. Cinnabar, e.g., has three formula units per unit cell~\cite{Schleid1999} whereas metacinnabar only has one. This fact is responsible for a much larger number of ir and Raman active phonons in cinnabar than in metacinnabar. Their frequencies are investigated here. Cinnabar has two chiral (enantiomorphic) modifications (space groups no. 152 (D$_3^4$) and no. 154 (D$_3^6$), primitive cell composed of two coaxial helices, one with three S atoms, the other with three Hg atoms). These modifications rotate the plane of polarization of light propagating along the $c$-axis in opposite directions (optical activity). They also give rise to linear \textbf{k} terms in the dispersion relations of phonons and electrons, the latter even in the absence of spin-orbit (SO) coupling (linear terms appear in $\beta$-HgS only when SO is taken into account). In the present work we have measured the specific heat of natural and synthetic crystals of $\alpha$-HgS with various isotopic compositions and compared the results with \textit{ab initio} calculations based on the electronic band structures mentioned above. Because of the uniaxial crystal structure, there are two different spectra of the dielectric functions, depending on whether the electric field is polarized parallel or perpendicular to the $c$-axis. We have calculated the corresponding electronic spectra and compared them with unpublished experimental data by Zallen \textit{et al.}.~\cite{Zallen2010}

This work is organized as follows:

Section \ref{SectionTheory} discusses the details of the theoretical calculations, which use the VASP~\cite{Kresse1996}   and the ABINIT set of   codes.~\cite{Gonze2002}
Section \ref{SectionExperimental} provides necessary experimental details.
Section \ref{SectionBand} presents the calculated \textit{ab initio} band structures (with emphasis on SO  and
linear $\textbf{k}$ effects) and the lattice parameters (including the bulk modulus $B_0$ ) obtained by optimization of the total enthalpy.

Section \ref{SectionOptical} displays the electronic dielectric functions obtained from the ABINIT-calculated band structures and compares them with the results obtained by Kramers-Kronig analysis of the reflectivity spectra   measured by Zallen \textit{et al.}.\cite{Zallen2010}

Section \ref{SectionPhonon} is devoted to the \textit{ab-initio} calculation of the phonon dispersion curves and a comparison of the frequencies obtained at the $\Gamma$-point of the BZ with Raman and ir measurements.
Section \ref{SectionHeat} discusses the results of our heat capacity measurements and finally
Section \ref{SectionConclus}  is devoted to the conclusions.

\section{Theoretical Details}\label{SectionTheory}

The calculations reported here have been performed with two different
implementations of Density Functional Theory (DFT), VASP and ABINIT:
The Vienna simulation package VASP (see Refs.~\onlinecite{vasp1,vasp2} and references therein)  performs \textit{ab initio} electronic structure calculations with the pseudopotential method and with the spin-polarized density functional theory including a self-consistent treatment of SO coupling. The set of plane waves employed extended up to a kinetic energy cutoff of 370 eV for $\alpha$-HgS.
Such a large cutoff was required to achieve highly converged results within the projector-augmented wave (PAW) scheme.~\cite{paw1,paw2}
The semicore 5$d$ and 6$s$ electrons of Hg were included explicitly in the calculations. The PAW method takes into account the full nodal character of the all electron charge density distribution in the core region.  The exchange-correlation energy was initially taken in the local density approximation (LDA) with the Ceperley Alder prescription~\cite{CA}  of the exchange-correlation energy, and we also employed the  generalized gradient approximation, GGA, with the PBE prescription.~\cite{pbe} We used a dense Monkhorst-Pack grid  for Brillouin zone (BZ) integrations in order to assure highly converged results  to about 1-2 meV per formula unit. We also utilized an accurate procedure  in the calculations in order to obtain very well converged forces which were used for the calculation of the dynamical matrix. At each selected volume, the structures were fully relaxed to their equilibrium configuration through the calculation of the forces on atoms and the stress tensor~\cite{rmp}. In the relaxed equilibrium configuration, the forces are less than 0.004 eV/\AA\ and the deviation of the stress tensor from a diagonal hydrostatic form is less than 1 kbar (0.1 GPa).

In the ABINIT implementation\cite{Gonze2009,Gonze2005}, only the valence electrons are taken into account by using the
Hartwigsen-Goedecker-Hutter relativistic separable dual-space
pseudopotentials\cite{Hartwigsen1998} with an energy cutoff of 60 Ha. A detailed
testing
was performed to ensure convergence at the chosen cutoff. A Local
Density Approximation (LDA) was used for the exchange-correlation Hamiltonian, but some
tests were also performed with the PBE-GGA exchange correlation Hamiltonian\cite{Perdew1996}; only
small differences were found.
A grid of 6$\times$6$\times$4 was used to describe electronic and vibrational properties of
$\alpha$-HgS in both symmetry groups.

Several tests were
performed with finer grids to check our results. After convergence was
reached, an inner
stress within the unit cell  of 1$\times$10$^{-4}$
GPa remained.

Geometrical relaxation was performed up to forces less than 3x10$^{-4}$
eV/\AA. The ABINIT code was used to obtain the  optical and the
vibrational properties. In the latter case we
used the response function implementation.\cite{Gonze1997a,Gonze1997b} The dynamical matrices
were calculated for a grid of 6$\times$6$\times$3 and four different grid shifts, with a
total of 83 matrices (including the $\Gamma$ point). After obtaining the matrices, a Fourier
interpolation was used, as implemented in the set of ABINIT codes (anaddb)
described in Ref. \onlinecite{Baroni2001}, to increase the density of \textbf{q} points. However,
in order to calculate the optical properties we used the ABINIT
implementation in two different versions: (a) calculation of the
electronic interband transitions, as described in Ref. \onlinecite{Sharma2004} and (b) a different
approach, where an inversion of the dielectric
matrix is performed as described in Refs.~\onlinecite{Baroni1986,Giantomassi2009}.

\section{Experimental}\label{SectionExperimental}

Natural cinnabar crystals on dolomite were purchased from a mineral dealer (Sebastian Barzel, Ganghoferstr. 1, Berlin. Origin: China). A chemical microanalysis performed on a set of selected crystals gave an Fe content of less than 50 at-ppm Fe and a ratio of Hg and S of 1:0.994(4), very close to the ideal stoichiometry.
Artificial cinnabar crystals were grown from $\alpha$-HgS (Chempur, 99+\%) by chemical vapor  transport using iodine as transport agent ($\sim$1 mg\,cm$^{-3}$) in a temperature gradient of 340$^{\rm o}$C and 180$^{\rm o}$C.~\cite{Schafer1962}
This procedure lead to the simultaneous deposition of the red and the black modification. Well shaped crystals of the red cinnabar were carefully separated and selected from the growth product.
$^{34}$S (enrichment 99.6\%) was purchased from (Trace Sciences International Corp., 40 Vogell Road, Unit 42, Richmond Hill, ON) and reacted in the gas phase at 300$^{\rm o}$C with elementary Hg (Chempur, 99,999+\%) in a ratio 1:1. The resulting red cinnabar was subsequently refined by a gas phase transport reaction similar to that described above.
Isotope enriched Hg (isotope composition
$^{196}$Hg: ${3.43}$\%,  $^{198}$Hg: ${43.0}$\%, $^{199}$Hg: ${30.1}$\%, $^{200}$Hg: ${8.6}$\%, $^{201}$Hg: ${4.39}$\%, $^{202}$Hg: ${8.48}$\%, $^{204}$Hg: ${1.97}$\%) was obtained as HgO (also from Trace Sciences International Corp.) and reacted with S at 400$^{\rm o}$C. The resulting black modification of HgS was subsequently refined,  recrystallized and converted into $\alpha$-HgS by a gas phase transport reaction as described before.
The heat capacities were measured on samples of typically $\sim$20 mg between 2 and 350 K
with a physical property measurement system (Quantum Design,
6325 Lusk Boulevard, San Diego, CA) as described in
detail in Ref.~\onlinecite{Serrano2006}. Unfortunately, for the Hg isotope enriched sample we could only reliably separate  several smaller crystals the total mass of which did not exceed $\sim$2.5 mg. The bulk of the Hg isotope enriched sample consisted of black material, presumably mostly $\beta$-HgS.

\section{Electronic Band Structure}\label{SectionBand}

Some of our band structure calculations included SO interaction, others not. The latter (NOSO) were performed in order to examine the effect of the SO interaction (not always small because of the large SO coupling in the Hg atoms) and also to reduce computational time whenever needed (i.e., for the calculation of optical and phonon spectra).

\begin{figure}[htp]
\includegraphics[width=8cm ]{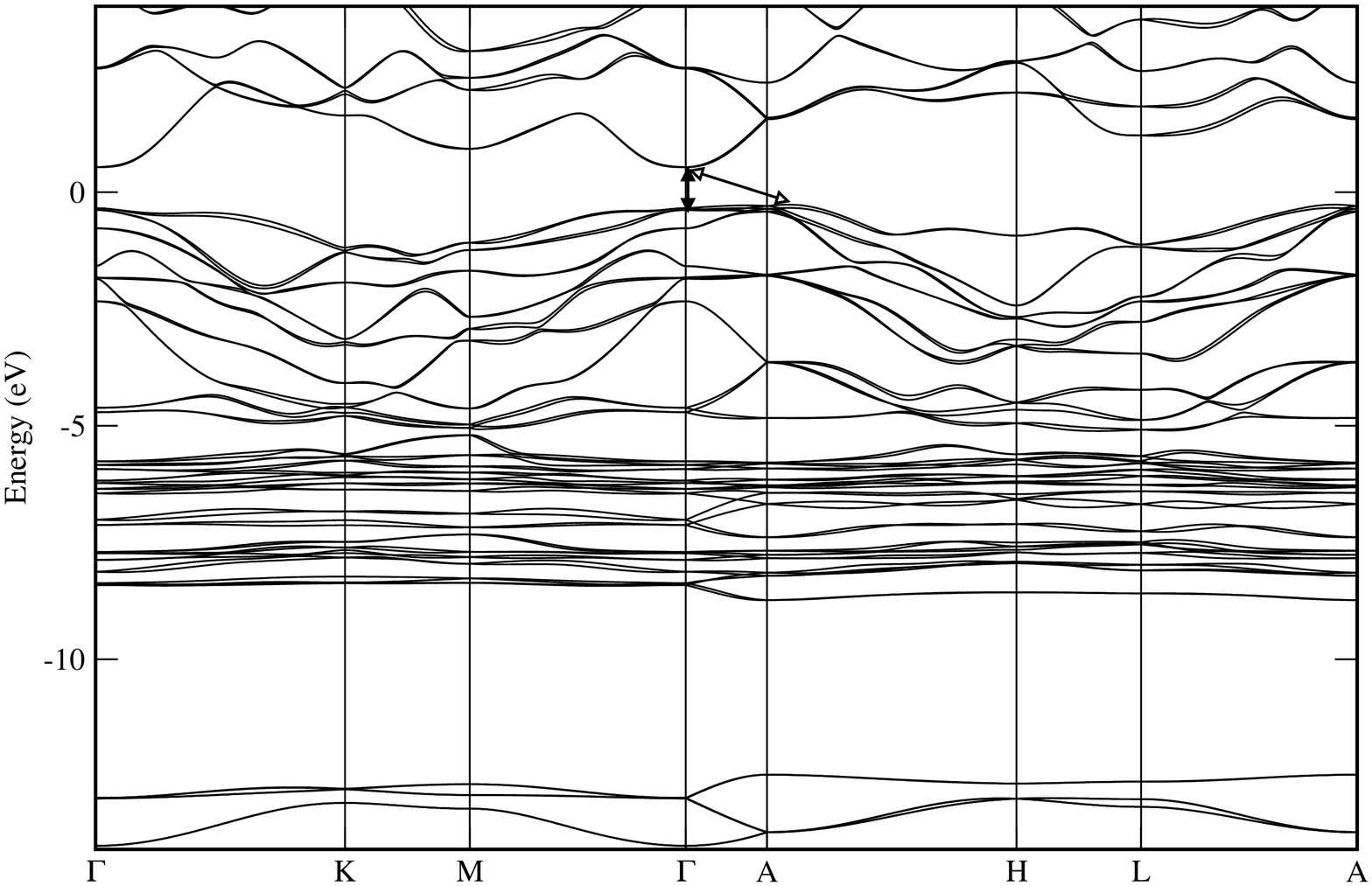}
\caption{Electronic band structure (VASP LDA SO) of $\alpha$-HgS. The arrows near $\Gamma$ are discussed in Fig. \ref{Fig2}.} \label{Fig1}
\end{figure}

\begin{table*}[htp]
\begin{center}
\begin{tabular}{ l  c  c  c  c  c c  c } \hline\hline
&  $V_0$  (\AA$^3$)  &    $a$ (\AA) &    $c$ (\AA) & $u$      & $v$ &   $B_0$ (GPa) & $B_0'$ \\ \hline
VASP-LDA NOSO       &	130.40 &	4.024   &	9.299  & 0.68    &	0.47 & - & - \\
VASP-LDA SO &	130.00 &	4.023   &	9.274  & 0.68    &	0.48	& 42.89 &	 6.25\\
ABINIT-LDA
NOSO &	131.265	& 4.0415 &	9.280 &	0.68 &	0.47 & 35 & 10 \\
ABINIT-LDA
SO &	139.92 &	4.0382 &	9.9081 &	0.675	 & 0.47 & 34.55 & 11.1\\
VASP-GGA
NOSO &	164.99	 & 4.408 &	9.746	& 0.75	& 0.47 & - & -\\
WIEN2K-GGA
NOSO$^a$  &	154 &	4.29 &	9.50 &	0.72 &	0.50 &	22.5 &	4.8  \\
measured &	141.54$^b$ & 	4.15$^b$  &	9.50$^b$ &	0.720$^b$  &	0.492$^b$  &	37$^c$  &	4$^c$ \\
\hline\hline
$^a$all data from Ref.~\onlinecite{Sun2005}\\
$^b$Ref.~\onlinecite{Schleid1999}\\
$^c$Ref.~\onlinecite{Fan2009}\\
\end{tabular}
\end{center}
\caption[]{Cell volume, lattice parameters, $u$ and $v$ and bulk moduli of $\alpha$-HgS  resulting from calculations compared to experimental data.}
\label{RKKTab1}
\end{table*}

\begin{figure}[htp]
\includegraphics[width=8cm ]{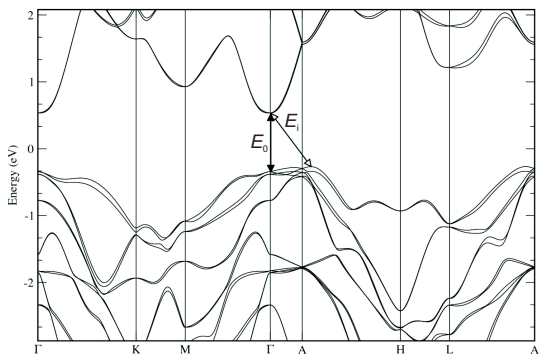}
\caption{Electronic band structure (VASP LDA SO) of $\alpha$-HgS in an enlarged energy scale. $E_0$ and $E_{\rm i}$ mark the direct and the indirect gap, respectively.} \label{Fig2}
\end{figure}

We show in Fig. \ref{Fig1} the  electronic band structure of $\alpha$-HgS in an extended  energy range along several high symmetry directions of the BZ (the BZ and the notation for the high symmetry points can be seen in Ref.~\onlinecite{Doni1979} ) including SO interaction, although in this figure its effects can only be clearly observed for the 30 bands derived from Hg 5$d$ states ( -5 to -8 eV range). Beside these bands, the figure includes bands in which the 3$s$ electrons of sulfur are dominant (-12 to -14 eV), sulfur 3$p$ valence bands (-5 to 0 eV) and empty conduction bands in which Hg 6$s$ and 6$p$ states dominate. The calculation was performed with the VASP code, using the LDA to approximate exchange and correlation. Similar results were obtained with the ABINIT code and thus will not be presented here. Calculations, performed with the WIEN2K code using the generalized gradient approximation (GGA) for exchange and correlation have been already published.~\cite{Sun2006}  They were used in Ref. \onlinecite{Sun2006} to investigate the pressure dependence of the band structure and the phase transition from cinnabar to the rock salt structure (at about 25 GPa~\cite{Sun2005}). Figure \ref{Fig1} and \ref{Fig2} show an indirect valence-to-conduction energy gap at 0.82 eV (from near A to the $\Gamma$ point of the BZ) followed by a direct one at 0.89 eV. These gaps are considerably smaller than the experimentally reported ones ($\sim$ 2.25 eV~\cite{Zallen1967}). The difference is related to the so-called \textit{gap problem} which appears when local density functionals are used in the calculations. It is semiempirically removed by \textit{ad hoc} adjusting the gap with a \textit{scissors operator}.~\cite{Doni1979} We present in Figs. \ref{Fig1} and \ref{Fig2} raw data obtained using the LDA. For the calculation of optical spectra we shall add 1 eV as a scissors operator, a correction determined  for $\beta$-HgS by Svane~\cite{Svane2010,Cardona2009} with the GW Hamiltonian. The Hg 5$d$ bands (cf.  Fig. \ref{Fig1}) cluster in two groups centered between 6 and 7.8 eV below the top of the valence bands (TOV), the splitting resulting from a combination of the orbital field and the SO interaction. Correspondingly, two peaks are seen in the photoemission spectra centered at 7.0 and 8.8 eV.~\cite{Shevchik1973}  While the width of these bands agrees with the calculated one, their higher binding energy (by about $\sim$1 eV) may be due to an LDA effect similar to that which causes the gap problem.~\cite{Christensen2009} In Ref. ~\onlinecite{Christensen2009} this energy difference was calculated to be 1.2 eV for wurtzite InN.
The sulfur 3$s$ bands are centered around 12.5 eV. They probably correspond to a peak observed in photoemission at the same energy (see Fig. 7 of Ref. ~\onlinecite{Shevchik1973}).

In order to better see the indirect ($E_{\rm i}$ ) and direct ($E_{\rm 0}$) gaps, as well as the SO splitting of the TOV, we show in Fig. \ref{Fig2} the bands of Fig. \ref{Fig1} in an expanded energy scale. A lowest indirect gap $E_i$ = 0.82 eV is observed in this figure, followed by the already mentioned direct one at 0.89 eV. The SO splitting at $\Gamma$ is found to be $|\Delta_0|$ = 39 meV. An interesting question is the sign of this splitting in view of the fact that, anomalously, that of $\beta$-HgS has been found to be negative because of admixture of Hg 5$d$ electrons to the dominant S 3$p$ electrons. Depending on the amount of admixture, the splitting, which is rather small anyhow, can become either positive or negative.~\cite{Cardona2000}
We shall present evidence that the spin-orbit splitting $\Delta_0$ of $\alpha$-HgS is positive, contrary to that of  $\beta$-HgS.

\begin{figure}[htp]
\includegraphics[width=8cm ]{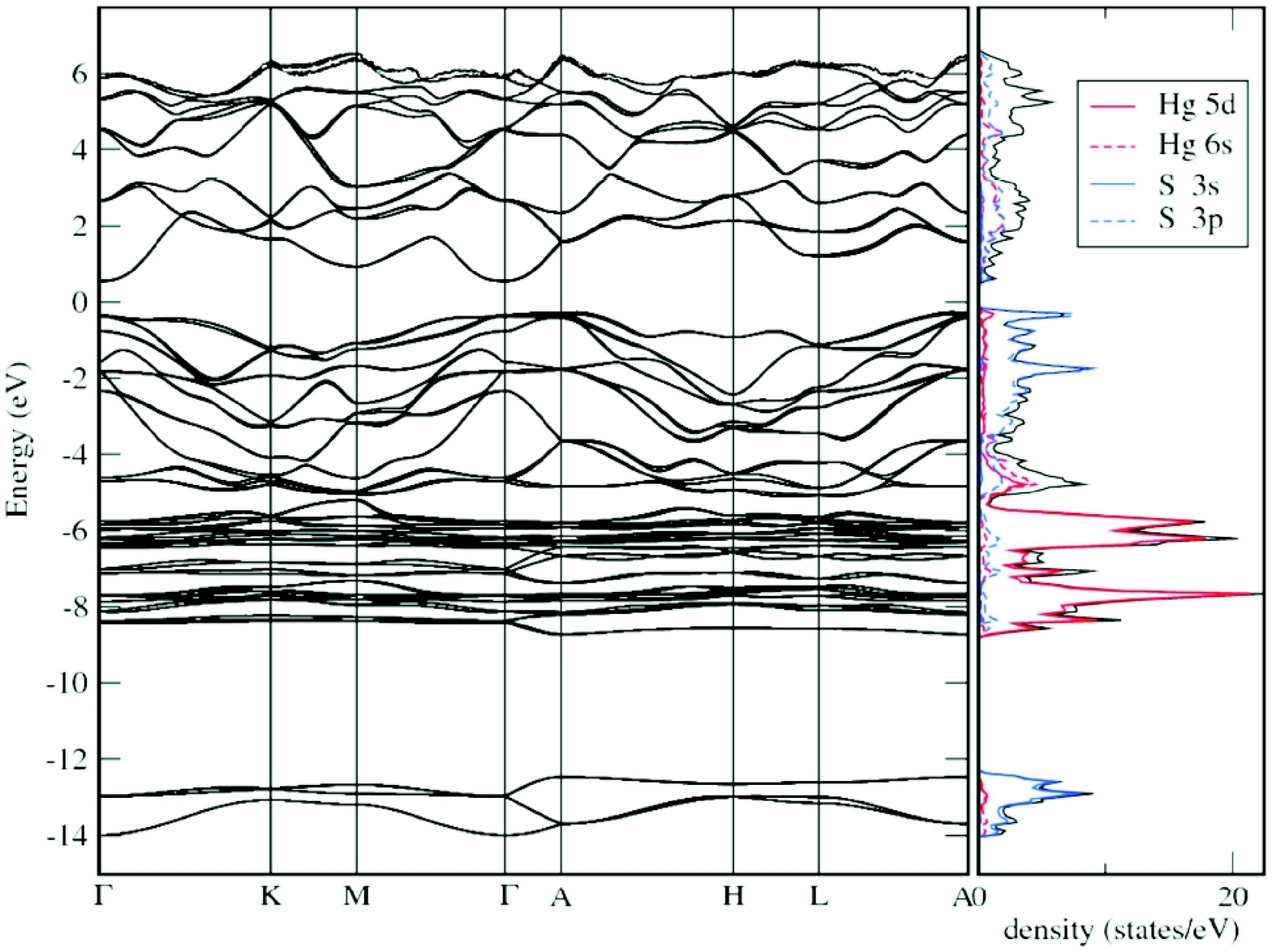}
\caption{(Color online) Electronic band structure (VASP LDA SO) of $\alpha$-HgS with partial electronic DOS as labeled in the inset of the r.h.s. of the figure.} \label{Fig3}
\end{figure}

By downloading the eigenvectors obtained from the VASP code, with and without SO coupling, we have obtained the symmetries of the corresponding states around the direct gap (at $\Gamma$). The symmetries corresponding to the simple $D_3$ point group (without spin) are two singlets ($\Gamma_1$ and $\Gamma_2$ ) and a doublet $\Gamma_3$.~\cite{Tinkham1964}    The double group (including spin) has three representations with symmetries $\Gamma_6$ (a doublet),  and two singlets ($\Gamma_4$, $\Gamma_5$), degenerate on account of time-reversal.\cite{Koster1957}  The symmetry of a single spin is $\Gamma_6$. We found the state at the bottom of the conduction band (Fig. \ref{Fig2}) to have $\Gamma_1$ symmetry, mainly consisting on each atom mainly of 6$s$ and 6$p_z$) atomic functions of Hg (see Fig. \ref{Fig3}). Its symmetry becomes $\Gamma_6$  when SO coupling is taken into account. The top of the valence band at $\Gamma$ is an orbital doublet ( $\Gamma_3$ symmetry), composed mainly of (3$p_x$, 3$p_y$) functions of sulfur, with some admixture of 5$d$ functions of Hg. The latter are responsible for the small (in $\beta$-HgS negative) value of $\Delta_0$. The next lowest $\Gamma$ state (0.4 eV below $\Gamma_3$) has $\Gamma_2$ orbital symmetry. Under SO interaction, the $\Gamma_3$ state splits into two spin doublets, the upper one having ($\Gamma_4$, $\Gamma_5$) symmetry, the lower $\Gamma_6$ . This ordering corresponds to a positive SO splitting, in which for the upper state the angular momenta of the spin and the orbit are parallel (see Fig. 5.1 of Ref.~\onlinecite{Cohen1989}) Hence these calculations suggest that $\Delta_0$ is positive for $\alpha$-HgS.

\begin{figure}[htp]
\includegraphics[width=7cm ]{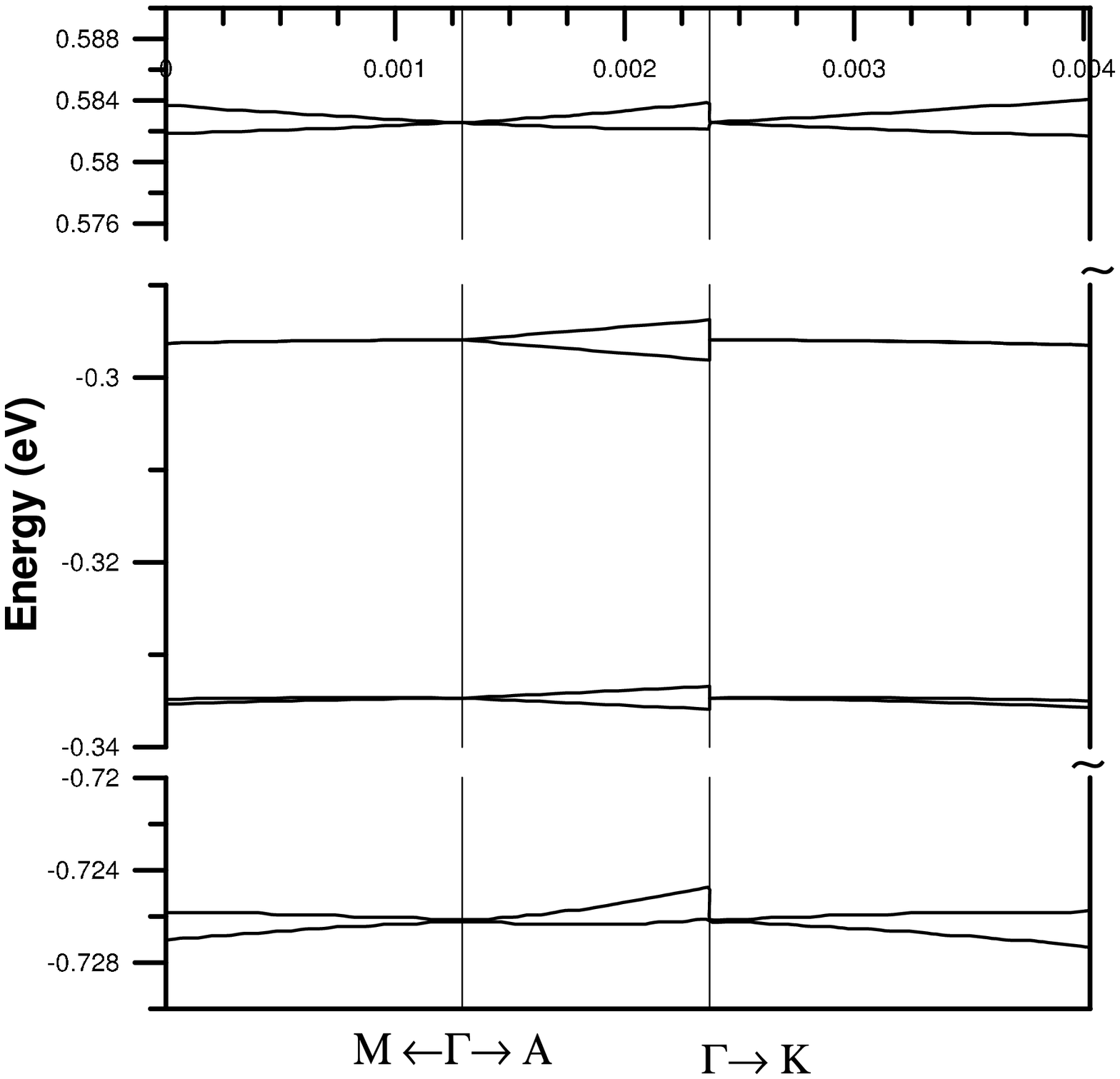}
\caption{Magnification of the electronic bands showing the linear splitting of the lowest conduction and the highest valence band near $\Gamma$ (in units of 2$\pi$$\times$\AA$^{-1}$).} \label{Fig4}
\end{figure}

Time reversal compels all double group states to be doubly degenerate at $\Gamma$. This degeneracy, however can split away from $\Gamma$, resulting in the so-called linear $\Gamma$ terms which we have already discussed for  $\beta$-HgS (zb structure).\cite{Cardona2009} While in $\beta$-HgS only $\Gamma_8$ quadruplets exhibit linear in \textbf{k} splittings,   most states (not all) show these splittings for the $\alpha$-HgS structure. Most interesting is the splitting of the $\Gamma_3$ orbital doublets for \textbf{k} along the trigonal axes ($\Gamma$ - A in Figs. \ref{Fig2} and \ref{Fig3}). This splitting is related to the chirality of the structure and its optical activity: it reverses its sign when going from the space group no. 152 ($D_3^4$) to the space group no. 154 ($D_3^6$) structure, as does its optical rotatory power.~\cite{Langlois1973} The structure of the linear in \textbf{k} splittings can be investigated  using so called invariants, i.e. Hamiltonian-like functions linear in $k$ which have the full symmetry of the crystal.~\cite{Bir1972}  Such invariant is, for the symmetry at hand:

\begin{equation}
H_k = C_z J_z k_z + C_x(J_x\,k_x + J_y\,k_y),
\label{Eq1}
\end{equation}

where the coefficients $C$ depend on the state under consideration and $J$ transforms like angular momentum ${J}$ = 1  under the symmetry operations of  $D_3$ . In the absence of spin, the expectation values of Eq. (\ref{Eq1}) vanish for states of symmetry $\Gamma_1$, $\Gamma_2$, $\Gamma_4$ , and  $\Gamma_5$ . As already surmised, the only coefficient which does not vanish in the absence of spin is  $C_z$. We have calculated for the top of the valence band in the absence of spin $C_z$ = 0.22 eV\,\AA.
When spin is included, all states of $\alpha$-HgS  exhibit splittings linear in k except those of ($\Gamma_4$, $\Gamma_5$) symmetry for \textbf{k} along A-M. This is shown in Fig. \ref{Fig4} for the lowest conduction band and the 3 highest valence bands at $\Gamma$. The calculated values of the coefficients $C_k$  (the splitting is defined as ($\pm$) $C_k\cdot k$) are displayed in Table \ref{RKKTab3}.  Note that for k along $\Gamma$-M and $\Gamma$-K $C_k$ is small but not exactly zero for the highest valence band. This is probably a computational error resulting from having had to deal with very small energy differences.

It is easy to understand this vanishing of $C_k$ for \textbf{k} along $\Gamma$-M and $\Gamma$-K on the basis of Eq. (\ref{Eq1}). The $\Gamma_4$ and $\Gamma_5$ doublet has the same transformation properties as $J_z$=3/2 and $J_z$=-3/2 angular momentum states. $J_x$  and $J_y$  can only connect states which differ in angular momentum $J_z$ by $\pm$1, hence there can be no linear splittings of the $\Gamma_4$ , $\Gamma_5$  states for \textbf{k} along either $k_x$ or $k_y$.  This observation, represented also in Fig. \ref{Fig4}, indicates that the ($\Gamma_4$, $\Gamma_5$ ) states correspond to the top of the valence band and confirms the positive sign attributed to $\Delta_0$ above. Notice that the non-zero values of Table \ref{RKKTab2}, especially those for \textbf{k} along the $\Gamma$-A direction, are similar to those found for $\beta$-HgS.\cite{Cardona2009} However, while a sign was attached to the values in Table \ref{RKKTab2} of Ref. \onlinecite{Cardona2009}, corresponding to placing the anion at the origin of coordinates~\cite{Koster1957}, the intricacies of  $\alpha$-HgS and the corresponding VASP program did not allow us to make this assignment in Table \ref{RKKTab2} with respect to the two possible enanthiomorphic varieties (the signs for space group  no. 152 variety should be opposite to those for no. 154). This sign should determine the corresponding sign of the dispersion of the optical activity, which has hitherto not been determined.~\cite{Langlois1973}

\begin{table*}[htp]
\begin{center}
\begin{tabular}{ l c c c c c c   } \hline\hline
                          & $\Delta_0$ & $d \Delta_0/d P$ & $d ln \Delta_0/d lnV$ & $E_0$ & $d E_0/d P$ & $d E_0 /d lnV$ \\
                          & (meV)   & (meV/GPa) &         & (meV) & (meV/GPa)  & (meV)\\ \hline
$\alpha$-HgS VASP-LDA     & 39	    &      5.1 &	0.52  &   890 & -97.7 &  4661.4\\
$\beta$-HgS VASP-LDA      & -111	&    -3.4	&   0.234 &  -573 & 17.4 & -134.7 \\
$\alpha$-HgS ABINIT-LDA   & -       & -         &        -&     - &  -& - \\
$\beta$-HgS ABINIT-LDA    & -180    &  -        &        -& - 580 & - & - \\
 \hline\hline
\end{tabular}
\end{center}
\caption[]{Spin-orbit splitting $\Delta_0$ at the top of the valence band of cinnabar ($\alpha$-HgS) and
metacinnabar ($\beta$-HgS) and its derivatives vs. pressure and volume (the latter logarithmic).
Also, direct gap at the $\Gamma$-point of the BZ (not corrected for the "gap problem") together with the corresponding pressure and volume derivatives.}
\label{RKKTab2}
\end{table*}

\begin{table}[htp]
\begin{center}
\begin{tabular}{ l  c  c  c } \hline\hline
energy  & $C$ ($\Gamma$ - A) & $C$ ($\Gamma$ - M) &  $C$ ($\Gamma$ - K)  \\
(ev) & (eV$\times$\AA) & (eV$\times$\AA) & (eV$\times$\AA) \\ \hline
0.62 (cb) & 0.132       & 0.012 & 0.120 \\
-0.28 (vb) & 0.322      & 0.002 & 0.006 \\
-0.32 (vb) & 0.182      & 0.033 & 0.036 \\
-0.70 (vb) & 0.092  & 0.068 & 0.074 \\
\hline\hline
\end{tabular}
\end{center}
\caption[]{
Coefficients of the terms linear in k for the states around the direct gap of cinnabar. The linear splitting is $\pm C$. These splittings were calculated including spin-orbit interaction. Because of the chirality, linear terms in $k$ also appear in the absence of spin-orbit interaction for orbital states of $\Gamma_3$ symmetry such as those at the top of the valence band. In this case $C$ = 1.357/2$\pi$ = 0.216 eV$\times$\AA.}
\label{RKKTab3}
\end{table}

\section{Optical Properties}\label{SectionOptical}

\begin{figure}[htp]
\includegraphics[width=8cm ]{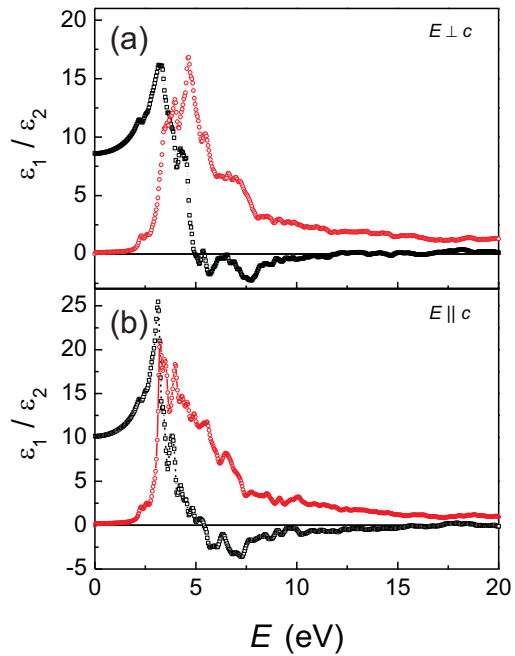}
\caption{(Color online) Calculated  spectra (ABINIT LDA, NOSO) of $\epsilon_{1,2}$ ({\tiny{$\square$}} and  {\color{red} $\circ$}, respectively) for electrical field (a) $E \perp c$ and (b) $E \parallel c$.} \label{Fig5}
\end{figure}

We display in Fig. \ref{Fig5}  the calculated real and imaginary parts of  the dielectric function for the electric field perpendicular and parallel to the trigonal axis for $\alpha$-HgS.

\begin{figure}[htp]
\includegraphics[width=8cm ]{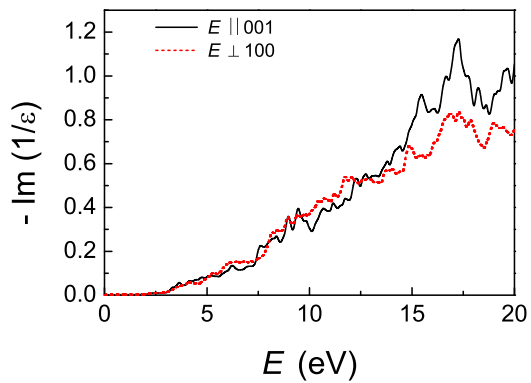}
\caption{(Color online) Calculated electron energy loss function  function of $\alpha$-HgS. } \label{Fig7}
\end{figure}

The calculations were performed using the ABINIT code without SO coupling and with the LDA Hamiltonian for exchange and correlation. We separated the calculated conduction and valence bands by a 1eV "scissors operator" which we took to be equal to that obtained with the GW approximation for $\beta$-HgS.\cite{Svane2010} A Lorentzian broadening of 0.1 eV was applied to all transition energies. Notice that $\epsilon_1$ becomes zero for $E\approx$17 eV, an energy which corresponds to the plasma frequency of the valence electrons, with some contribution of the 6$d$ semi-core electrons of Hg. This plasma frequency can be obtained more accurately from a plot vs. energy of the electron energy loss (EEL) function

\begin{equation}
F(EEL) = -\Im(1/\epsilon),
\label{Eq2}
\end{equation}

\noindent
as shown in Fig. \ref{Fig7}. Both EEL functions, for $E$ parallel and perpendicular to the $c$-axis, show a broad band peaking around  17.5 eV.

The analysis of the overall optical response in the 0-20 eV range can be completed by calculating the effective number of electrons contributing up to the energy $E$, using the expression\cite{Ehrenreich1966}

\begin{equation}
N_{\rm{eff}} = \frac{2}{\pi}\frac{m_e}{4\pi e^2}\int_0^E\!\epsilon_2(E')E'dE'
\label{Eq3}
\end{equation}

\begin{figure}[htp]
\includegraphics[width=8cm ]{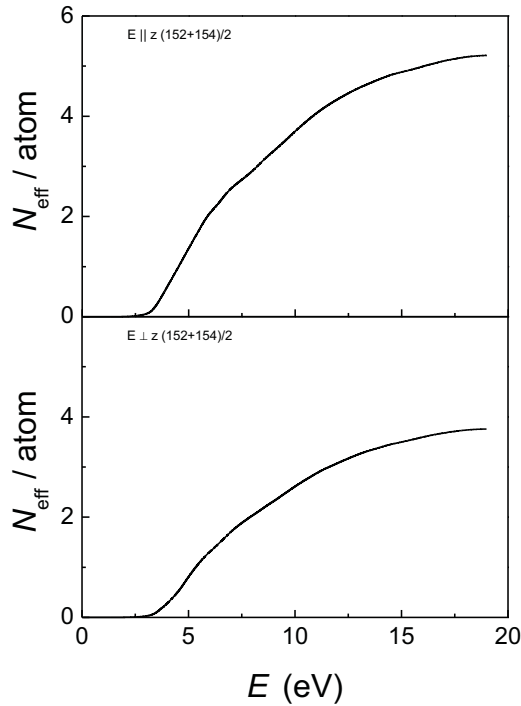}
\caption{Effective number of electrons per atom $N_{\rm{eff}}$  contributing to the optical absorption up to the energy $E$, as obtained by integrating the  $\epsilon_i$ of Fig. \ref{Fig5} with Eq. (\ref{Eq1}).} \label{Fig5b}
\end{figure}

The energy dependence of  $N_{\rm{eff}}$  obtained by substituting into Eq. (\ref{Eq3}) the values of $\epsilon_{\rm i}(E)$ given in Fig. \ref{Fig5}  are shown in Fig. \ref{Fig5b}. According to this figure, $N_{\rm{eff}}$  has not attained saturation at the maximum energy of $\approx$19 eV. The values reached at this energy are 5 electrons per atom for the parallel component and 4 for the perpendicular, rather close to the average number of valence $s$ and $p$ electrons per atom. Hence we conclude that the 5$d$ electrons of mercury contribute to $N_{\rm{eff}}$ mainly at much higher energies.\cite{Cardona1970} It is interesting to
note that up to 19 eV  $N_{\rm{eff}}$  is larger for the parallel (5 electrons/atom) than for the perpendicular (4 eV/atom). It is not easy to find a simple reason for this fact, which is also reflected in the higher value of  $\epsilon_1(0)$ for $E||c$  (ordinary  rays) than for $E\perp c$ (extraordinary rays) see Fig. \ref{Fig5}.

\begin{figure}[htp]
\includegraphics[width=8cm ]{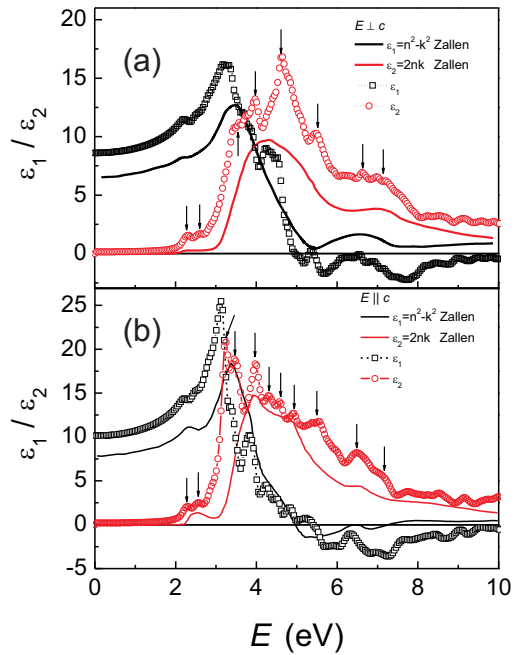}
\caption{(Color online) Calculated (ABINIT LDA, NOSO) ({\tiny{$\square$}} and $\circ$) and measured spectra (black and red solid lines) of $\epsilon_{1,2}$ ($E \perp c$) and $\epsilon_{1,2}$ ($E \parallel c$). The experimental data  have been taken from Zallen \textit{et al.} (Ref. \onlinecite{Zallen2010}). The vertical arrows mark structures in the calculated $\epsilon(\omega)$ curves. Their energies are listed in Table \ref{RKKTab4}.} \label{Fig6}
\end{figure}

The calculated spectra of Fig. \ref{Fig5} are compared in Fig \ref{Fig6} with unpublished experimental results of Zallen \textit{et al.} obtained by Kramers-Kronig analysis of normal incidence reflectance data.\cite{Zallen2010}  The experimental values are throughout the whole spectral range lower than the calculated ones, a fact which cannot be completely attributed to experimental errors:  measurements at energies below the gap ($E\approx$ 0) give $\epsilon_1$ = 8.4  for $E\parallel c$  and $\epsilon_1$ = 6.9 for  $E\perp c$    (note that $\epsilon_1$ = 8.4  is the highest value of $\epsilon_1(0)$ found for any mineral). However, although the calculations give values of $\epsilon$  larger than the experimental ones, the calculated birefringence ($\epsilon_{1 {\parallel}}$ - $\epsilon_{1 {\perp}}$) and its sign  agrees roughly  with the measured one.\cite{Ayrault1973}   The calculated  $\epsilon_1$  and  $\epsilon_2$  show considerable structure: we have indicated the most prominent peaks in $\epsilon_2$  by arrows. Most of this structure seems to correspond to features observed experimentally: we have listed in Table \ref{RKKTab4}  the calculated positions of peaks in $\epsilon_2$  and compared them with experimental ones. The calculated absorption starts at about 2 eV but this energy depends on the value of 1 eV, somewhat arbitrarily taken for the scissors operator.  According to Zallen\cite{Zallen1967} an indirect gap appears about 10 meV below the direct one. In Fig \ref{Fig2} such gap can indeed be seen about 40 meV below the direct one. The difference is not surprising as it may be related to the gap problem already mentioned. In any case, the direct-indirect energy differences are so small that even \textit{ab initio} calculations are not likely to reproduce them exactly. It is already remarkable that we can get its sign.
Notice in Fig. \ref{Fig2} of Ref.~\onlinecite{Zallen1967} that near the absorption edge the extraordinary rays ($E \perp c$)  are more strongly absorbed than the ordinary ones ($E \parallel c$). This is the opposite of what is shown in Fig. \ref{Fig5} for the overall absorption spectrum and its main peak at about 4.5 eV. Hence a crossover of the corresponding values of $\epsilon_2$  must take place at energies slightly above the absorption edge. Unfortunately in this region the calculations, as well as the extant measurements, are not accurate enough to reveal this fact.

\begin{table*}[htp]
\begin{center}
\begin{tabular}{ r c  c  c  c  c  c  c c c c c} \hline\hline
 theory $\epsilon_{2}$ ($E \perp c$)  &    2.28 & 2.60 & 3.55 & 3.98 & 4.61 & 5.51 & 6.64 & 7.14 & - & - & -\\
 exp $\epsilon_{2}$ ($E \perp c$) & 2.29 & - & 3.47 & 4.31 & 4.31 & 5.40 & 6.92 & 7.30 & - & - & -\\ \hline
 theory $\epsilon_{2}$ ($E \parallel c$) & 2.28 & 2.56 & 3.24 & 3.46 & 3.96 & 4.30 & 4.59 & 4.94 & 5.40 & 6.47 & 7.17\\
 exp $\epsilon_{2}$ ($E \parallel c$) & 2.35 & 2.55 & - & 3.56 & 3.96 & - & - & 4.94 & - & 6.61 & - \\
 \hline\hline
\end{tabular}
\end{center}
\caption[]{Comparison of calculated (see Figs. \ref{Fig6}(a) and (b)) and experimental energies (in eV) of peaks in $\epsilon_2(\omega)$ for $E \perp c$ and $E \parallel c$.~\cite{Zallen2010}}
\label{RKKTab4}
\end{table*}

\section{Phonon Properties}\label{SectionPhonon}

Regrettably there are no experimental data for the phonon dispersion relations throughout the whole Brillouin zone (BZ). Inelastic neutron scattering measurements are, in principle, possible. However, the rather large absorption cross section for the isotope $^{199}$Hg ($\sim$ 17\% abundance in the natural isotope mixture) would require an isotope enriched sample.
Given this problem, the most appropriate technique to map the whole BZ would be inelastic x-rays scattering. Although the resolution of this technique is rather limited ($\sim$10 cm$^{-1}$), small crystals ( $\sim$ 1 mm$^3$) can be handled. At this point, the only available experimental data are those obtained with Raman and ir spectroscopy.\cite{Zallen1970,Nusimovici1973}  In Ref. \onlinecite{Nusimovici1973} semiempirical calculations are presented based on a mixed valence-Coulomb force field adjusted so as to fit as well as possible the experimental results of Ref. \onlinecite{Zallen1970}, consisting of Raman and ir- active phonons at or very near the center ($\Gamma$) of the Brillouin zone. The 6 atoms per primitive cell give rise to 18 vibrational modes, 3 of which have zero frequency at $\Gamma$ (acoustic modes). Thus we are left with 15 modes, 5 $\Gamma_3$  doublets (ir and Raman active), 2 $\Gamma_1$ singlets (Raman active) and 3 $\Gamma_2$ singlets (ir active).~\cite{Nusimovici1973}  The ir active modes split into longitudinal and transverse, depending on whether the $E$-field is parallel or perpendicular to the scattering vector, the splitting being determined by Born effective charges.~\cite{Zallen1970}

\begin{table}[htp]
\begin{center}
\begin{tabular}{ c  c  c  c  c  } \hline\hline
symmetry & measured & semiemp$^a$ & NOSO &SO\\

 & (cm$^{-1}$) & (cm$^{-1}$) & (cm$^{-1}$) & (cm$^{-1}$) \\ \hline
$\Gamma_1$ & 43 & 43  & 39.1   & 39.2 \\
$\Gamma_1$ & 256 & 259  & 232.1   & 235.2 \\
$\Gamma_2^{\rm L}$ &   39  & 39 & 44.1 & 44.9 \\
$\Gamma_2^{\rm T}$ &   33  & 33 & 42.0 & 44.1 \\
$\Gamma_2^L$ &   141  & 146 & 159.6 & 158.4 \\
$\Gamma_2^{\rm T}$ &   110  & 109 & 158.7 & 158.6 \\
$\Gamma_2^{\rm L}$ &   361  & 363 & 337.7 & 339.1 \\
$\Gamma_2^{\rm T}$ &   333  & 338 & 325.6 & 327.7 \\
$\Gamma_3^{\rm L}$ &   48  & 30 & 44.1 & 44.1 \\
$\Gamma_3^{\rm T}$ &   43  & 1.7 & 42.0 & 42.0 \\
$\Gamma_3^{\rm L}$ &   91  & 86 & 87.9 & 88.4 \\
$\Gamma_3^{\rm T}$ &   87  & 88 & 83.1 & 83.3 \\
$\Gamma_3^{\rm L}$ &  108  & 109 & 122.4 & 120.7\\
$\Gamma_3^{\rm T}$ &  147 & 143 & 121.4 & 120.2\\
$\Gamma_3^{\rm L}$ &   288  & 299 & 267.2 & 269.3 \\
$\Gamma_3^{\rm T}$ &   280  & 286 & 259.3 & 261.7 \\
$\Gamma_3^{\rm L}$ &   350 & 349 & 323.5 & 326.2 \\
$\Gamma_3^{\rm T}$ &   342  & 339 & 319.2 & 323.5 \\
 \hline\hline
$^a$ Ref.~ \onlinecite{Nusimovici1973}
\end{tabular}
\end{center}
\caption[]{Phonon frequencies measured for $\alpha$-HgS. Also, semiempirical valence-Coulomb force field calculations by Nusimovici and Gorre.~\cite{Nusimovici1973} and the results of our \textit{ab initio} calculations with spin-orbit coupling (SO) and without spin-orbit-coupling (NOSO) are given for comparison.}
\label{RKKTab5}
\end{table}

The ratios of the L to T frequencies of the ir-active modes ($\Gamma_2$ and $\Gamma_3$) can be used to calculate the corresponding ratio of $\epsilon_0/\epsilon_{\infty}$  dielectric constants using the generalized Lyddane-Sachs-Teller (LST) relation:

\begin{equation}
\Pi_i (\frac{\omega^{\rm L}_{i}}{\omega^{\rm T}_{i}})^2 = \frac{\epsilon_0}{\epsilon_{\infty}}
\label{Eq4}
\end{equation}

\noindent
From the calculated NOSO frequencies in Table \ref{RKKTab5} we obtain using this LST relation:

($\epsilon_0/\epsilon_{\infty}$)$_\parallel$= 1.20 and
($\epsilon_0/\epsilon_{\infty}$)$_\perp$ =1.24. Using $\epsilon_{\infty}$ from Fig. \ref{Fig5} and the values of $\epsilon_0$ from the ABINIT program ($\epsilon_{0\parallel}$=12.39 and $\epsilon_{0\perp}$=10.33) we find the ratios ($\epsilon_0/\epsilon_{\infty}$)$_\parallel$=1.23 and ($\epsilon_0/\epsilon_{\infty}$)$_\perp$ =1.23 in good agreement with those obtained with the generalized LST relation.

We performed \textit{ab initio} calculations of the phonon frequencies vs. wavevector \textbf{k} using the ABINIT code within the LDA, as described in Section \ref{SectionTheory}. We had realized in past work that the calculated phonon frequencies could depend considerably on SO interaction in materials containing heavy ions (Bi, PbTe, HgTe).\cite{Diaz2007,Romero2008,Cardona2010}
Since mercury falls in this category, we performed a few calculations of phonon frequencies taking SO interaction and the relaxed lattice parameters (cf. Table \ref{RKKTab1}) into account. Table \ref{RKKTab5} summarizes the results of such calculations for the phonon frequencies at the $\Gamma$ point, together with their counterparts obtained without SO interaction (NOSO). For comparison,  experimental data from Raman and ir spectroscopy\cite{Zallen1970} and the results of semiempirical calculations fitted to the experimental data are also listed.\cite{Nusimovici1973}  The effect of SO interaction on the calculated frequencies is small, typically 1\% less than when it is not taken into account. Most of  the calculated frequencies become slightly closer to the measured ones than in NOSO calculation. Since the discrepancy between calculated and measured frequencies lies between 5 and 10\%, we decided, however, not to include SO interaction any further in the \textit{ab initio} calculations, so as to reduce computational time. The calculated LO-TO splittings of most modes are somewhat smaller than the measured ones, a fact that had already been observed in earlier work on zincblende and rocksalt type materials.\cite{Romero2008,Cardona2009} On the whole, the agreement between measurements and calculations is reasonable, especially when one considers the complicated crystal structure of  $\alpha$-HgS, with six atoms per unit cell.

\begin{center}
\begin{figure}[htp]
\includegraphics[width=7cm]{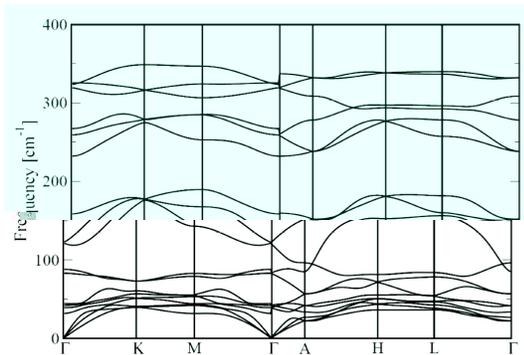}
\caption{Calculated (NOSO) phonon Dispersion of $\alpha$-HgS.} \label{Fig9}
\end{figure}
\end{center}

We display in Fig. \ref{Fig9} the phonon dispersion relations calculated with the NOSO restriction.  It is worth noticing in this figure that the five $\Gamma_3$ doublets at the $\Gamma$ point split linearly in \textbf{k} along the $\Gamma$-A direction, an effect similar to that  discussed in Section \ref{SectionBand} in the absence of SO interaction. These splittings, induced by the chirality of the structure, should be responsible for optical activity related to the ir-active phonons.

\begin{figure}[htp]
\includegraphics[width=8cm ]{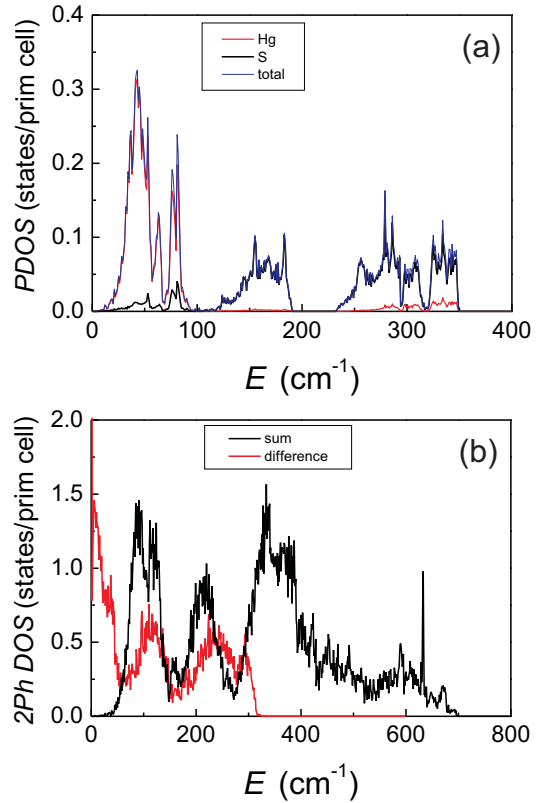}
\caption{(Color online) Phonon DOS including partial contributions, Hg and S, also two-phonon DOS for sums and differences of two phonons with equal \textbf{k}-vectors are shown.} \label{Fig10}
\end{figure}

We display in Fig. \ref{Fig10}(a) the phonon density of states (DOS) calculated from the dispersion relations of Fig. \ref{Fig9}, together with its decomposition in its S-like and Hg-like partial components. As expected, the low frequency band (0 - 100 cm$^{-1}$) corresponds mainly to Hg vibrations whereas that between 230 and 350 cm$^{-1}$ is mainly sulfur-like. It is interesting to notice that the intermediate band, between 110 and 190 cm$^{-1}$, is almost pure sulfur-like. We have plotted in Fig. \ref{Fig10}(b) the optical two-phonon DOS calculated with the restriction that the sum of the two participating phonon wave vectors be zero. Both DOS, corresponding to sums and differences of two phonons are given. These DOS will be of interest in the interpretation of carefully taken ir and Raman second order spectra, a rather delicate task in view of the large number of Raman and ir phonons active in first order. The difference spectra should vanish at low temperature according to the well known Bose-Einstein statistical factors. This, as well as the corresponding temperature dependence of the sum spectra, should help in identifying the second order Raman structures.

\section{Specific Heat}\label{SectionHeat}

\begin{figure}[htp]
\includegraphics[width=8cm ]{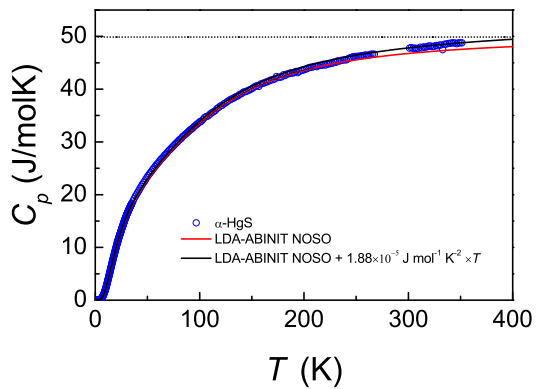}
\caption{(Color online) Heat Capacity of the artificial sample of $\alpha$-HgS ($\circ$). The results of our calculations ($C_v(T)$, NOSO) are represented by the (red) solid line. The (black) solid line shows $C_p(T)$ obtained by adding a term linear in temperature as described in detail in the text.
The (black) dotted vertical marks the Petit-Dulong limit of 6\,R, where $R$ is the molar gas constant.} \label{Fig11}
\end{figure}

Heat capacity measurements of $\alpha$-HgS have been performed before by  Khattak and coworkers in the temperature range up to $\sim$70K.~\cite{Khattak1981} In order to compare with our theoretical calculations we have extended these measurements up to $\sim$350K on a set of samples comprising natural and artificial crystals of $\alpha$-HgS. The results of these measurements are displayed  over the whole temperature range  in Fig. \ref{Fig11}.

\begin{figure}[htp]
\includegraphics[width=8cm ]{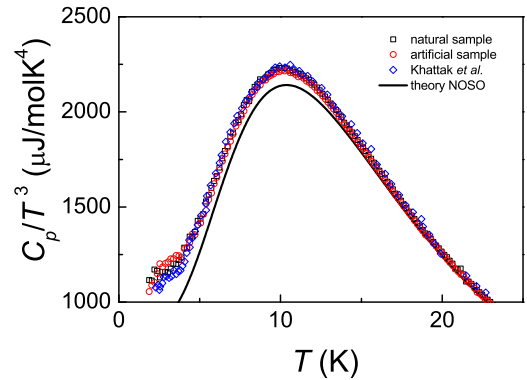}
\caption{(Color online) $C_p/T^3$ vs. $T$ representation of our heat capacity data. Natural $\alpha$-HgS sample: {\tiny{$\square$}}; artificial vapor phase grown sample: {\color{red} $\circ$}. The solid line represents our theoretical data (NOSO, for details see text). Earlier heat capacity data by Khattak \textit{et al.}\cite{Khattak1981} are represented by: {\color{blue}$\diamond$}.}
\label{Fig12}
\end{figure}

To emphasize the low temperature regime our experimental data
are also shown in the standard $C_p/T^3$ plot ('Debye plot') together with the data obtained by Khattak \textit{et al.} for comparison (Fig. \ref{Fig12}). Additionally, the results of our calculations based on the standard integration of the DOS of Fig. \ref{Fig10}(a) with the appropriate Bose-Einstein factors are also provided in both figures. Fig. \ref{Fig12} illustrates the very good agreement of our experimental data with those of Khattak \textit{et al.}. Furthermore, there is also no noticeable difference between our natural and artificial $\alpha$-HgS sample.

Figure \ref{Fig12}  reveals the characteristic maximum in the quantity $C_p/T^3$ at $\sim$10K which had actually been observed at 7K for the zincblende modification ($\beta$-HgS). The position of this maximum is usually determined by the lowest maximum of the phonon DOS which, according to Fig. \ref{Fig10}(a) takes place at 65K.  The ratio 65/10=6.5 is typical for the maximum of $C_p/T^3$  in many semiconductors.
Our theoretical data agree fairly well with the experimental data.
Near the maximum in $C_p/T^3$ theory falls short by about 3.5\%. Above 20 K, up to 250 K there is good agreement of experimental and theoretical results. Above 250 K the theoretical data approach the Petit-Dulong value of 6$\times R$, where $R$ is the molar gas constant, somewhat slower than the experimental data. This difference is due to
increasing importance of the difference of the
constant-volume and constant-pressure specific heats which is given by
\cite{Ashcroft}:

\begin{equation}
C_p(T) = C_v(T) + A(T) \cdot T,
 \label{Eq4}
\end{equation}

\noindent
where
\begin{equation}
A(T) = \alpha_v^2(T) \cdot B_0 \cdot V_{\rm{mol}}.
 \label{Eq5}
\end{equation}

\noindent
$\alpha_v(T)$ is the temperature dependent coefficient of the volume thermal
expansion, $B_0$ the bulk modulus and $V_{\rm mol}$ the molar volume
at $T$ = 0 K which amounts to 28.5 cm$^{3}$\,mol$^{-1}$.

Good agreement of the experimental and theoretical data is achieved if we choose for the coefficient  of the linear temperature increase an average value $A_{\rm 0}$ of

\noindent
\begin{equation*}
A(T) \approx A_{\rm 0}  = 1.88\times 10^{-5} {\rm{J mol^{-1} K^{-2}}}.
\end{equation*}

For an uniaxial system the volume coefficient of thermal expansion can be calculated from the linear thermal expansion coefficients, $\alpha_{\parallel}$ and $\alpha_{\perp}$, according to

\begin{equation}
\alpha_v(T) \approx \alpha_{\parallel}(293 K) + 2\cdot \alpha_{\perp}(293 K).
\label{Eq7}
\end{equation}

Using for the linear expansion coefficients the literature values for temperatures between 20$^{\rm o}$C and 200 $^{\rm o}$C,
$\alpha_{\parallel}$ = 18.8 $\times$10$^{-6}$ K$^{-1}$ and
$\alpha_{\perp}$ = 18.1 $\times$10$^{-6}$ K$^{-1}$ given by
Ohmiya~\cite{Expansion},
the experimentally determined linear increase given above  implies a bulk modulus, $B_0$, of

\begin{equation*}
B_0 = 41(1) {\rm{GPa}}
\end{equation*}

\noindent
in good agreement with the theoretical results (cf. Table \ref{RKKTab1}).

\begin{figure}[htp]
\includegraphics[width=8cm ]{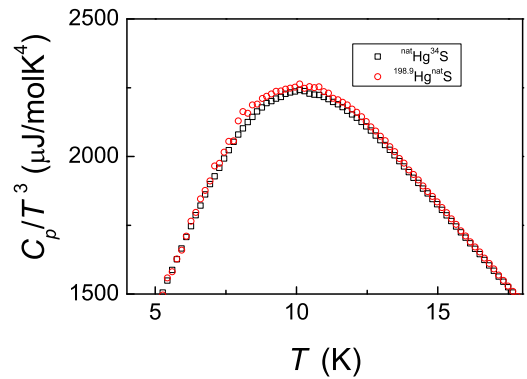}
\caption{(Color online)$C_p/T^3$ vs. $T$ representation of the heat capacities of $\alpha$-HgS samples grown from isotope enriched Hg and S as indicated in the inset.} \label{Fig13}
\end{figure}

\begin{figure}[htp]
\includegraphics[width=8cm ]{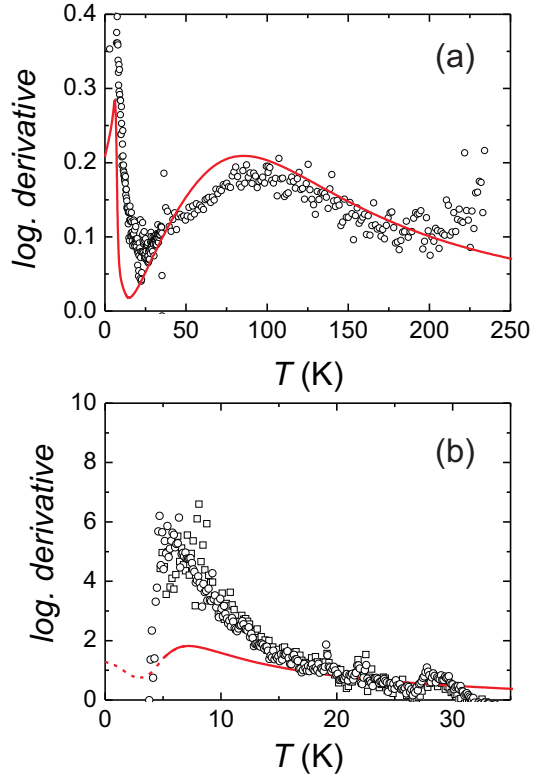}
\caption{(Color online) Logarithmic derivatives with respect to the masses $m$ of the constituents Hg or S,  d\,ln($C_p/T^3$)/d\,ln\,$m$ of the experimental heat capacities. (a) Compares $\alpha$-$^{\rm nat}$Hg$^{\rm nat}$S with a sample of  $\alpha$-$^{\rm nat}$Hg$^{34}$S. (b) Logarithmic derivative calculated by comparing samples of $\alpha$-$^{\rm nat}$Hg$^{\rm nat}$S and $\alpha$-$^{198.9}$Hg$^{\rm nat}$S   as described in detail in the text. Different symbols ($\circ$, {\tiny{$\square$}}) indicate independent runs on samples of the same preparation. The (red) solid lines represent the logarithmic derivatives obtained from our theoretical data calculated for various isotope masses.} \label{Fig14}
\end{figure}

Like in some of our previous work\cite{Romero2008,Cardona2010}  we have investigated the dependence of  $C_p/T^3$ on the isotopic masses of the two components of cinnabar and compared our experimental data with the theoretical results. Fig. \ref{Fig13} displays the low-temperature heat capacities of $\alpha$-$^{198.9}$Hg$^{nat}$S and $\alpha$-$^{nat}$Hg$^{34}$S. As expected, the heat capacity is slightly larger for the Hg isotopically enriched sample. Small differences in the heat capacity of the S isotope sample to the sample made from components with the natural isotope composition appear at higher temperature and are barely seen in the $C_p/T^3$ representation. They are, however, clearly revealed in the derivatives of $C_p/T^3$ versus to the isotope masses displayed in Fig. \ref{Fig14}.

Two pronounced features are observed in the logarithmic derivative versus the mass of the S constituent (Fig. \ref{Fig14}(a)), a rather pronounced sharp peak at $\sim$6 K and a broad hump with maximum at $\sim$100 K and probably a shoulder at $\sim$50 K. This broad feature reflects the two broad S-like bands  in the phonon DOS  (cf. Fig. \ref{Fig10} between 150 and 180 cm$^ {-1}$ and between 250 and 350 cm$^ {-1}$. The origin of the sharp low-temperature peak at 10 K is not immediately obvious from the phonon DOS.  We attribute it to a small mixture of Hg- and S-like phonons which also explains  the small spike in the
S-like phonon partial DOS at about 50 cm$^{-1}$.
All features, especially the broad hump peaking at $\sim$100 K, are  fairly well reproduced in position and magnitude by the results of our calculations.

The agreement between experiment (two separate runs on two separately prepared samples from the same batch of isotopically enriched Hg) and theory for the logarithmic derivative w.r.t. the mass of the Hg atoms (for the mass of the Hg constituent we took the mass-average of the Hg isotopes weighted by their abundance). A peak occurs in the logarithmic derivative at $\sim$6 K which corresponds well with the position expected from theory. But the experimental peak   overshoots the theoretical calculation by almost a factor of three. The origin of this difference might be due to a small traces of metacinnabar $\beta$-HgS or other related effects.  $\beta$-HgS has a significantly higher low-temperature heat capacity than $\alpha$-HgS, in a $C_p/T^3$ representation the peak is by a factor of three higher than that of $\alpha$-HgS.\cite{Cardona2009} Comparing the heat capacities of the two modifications of HgS we estimate that a 1.5\% admixture of $\beta$-HgS into our sample, probably hardly noticeable by visual inspection,  would be sufficient to explain the increase of the magnitude of the 6 K peak in the logarithmic derivative versus the Hg mass of $\alpha$-HgS. The high cost of mercury isotopes has prevented a more detailed analysis of this anomaly.

Finally, we demonstrate the relationship of the logarithmic derivatives versus temperature and versus the masses of the two  constituents following several previous results.\cite{Romero2008}

\begin{figure}[htp]
\includegraphics[width=8cm ]{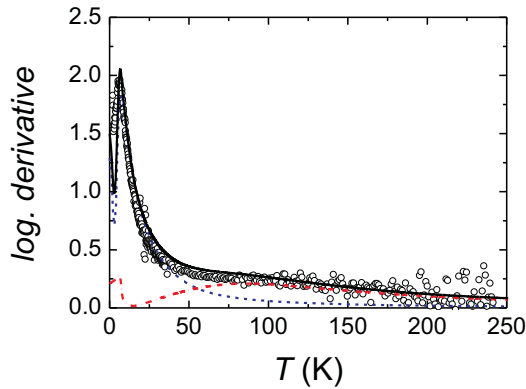}
\caption{Illustration of the relationship of the logarithmic derivatives w.r.t to temperature and the respective isotope masses given in Eq. \ref{Eq8}. The (blue) dashed and the (red) dashed-dotted lines represent the logarithmic derivatives versus the isotope masses obtained from our theoretical data for Hg and S, respectively. The (black) solid line is the sum of the two mass derivatives. $\circ$ denote the logarithmic derivative versus temperature of $\alpha$-$^{\rm nat}$Hg$^{\rm nat}$S.} \label{Fig15}
\end{figure}

For the relation of the temperature and mass derivatives in a two component system we have used the expression~\cite{Romero2008}

\begin{equation}
  \frac{1}{2}\,\,(3 + \frac{d
\ln (C_{p}/T^3)}{d \ln T})=\frac{d \ln (C_{p}/T^3)}{d \ln M_{\rm 1}} + \frac{d \ln (C_{p}/T^3)}{d \ln M_{\rm 2}}, \label{Eq8}
\end{equation}

where $M_1$ and $M_2$ are the masses of the two constituents, i.e. Hg and S, respectively, for $\alpha$-HgS.

This relation is nicely obeyed for $\alpha$-HgS, as is
is demonstrated in Fig. \ref{Fig15}.  Especially, the low temperature peak which emerges essentially from the logarithmic derivative versus the mass of Hg is very well reproduced, thus proving that the differences between experiment and theory discussed above originate from the experimental data.

\section{Conclusions}\label{SectionConclus}

\textit{Ab initio} electronic band structure techniques, especially those which use available computer codes such as VASP and ABINIT, are powerful methods to investigate electronic, optical, vibronic, and thermodynamic properties of crystals and compare the results with experimental data. Here we apply these techniques to cinnabar ($\alpha$-HgS) which has a chiral structure (space groups $D_3^4$ and $D_3^6$, three molecules per primitive cell)  more complicated than those usually dealt with. We calculate the electronic band structure and the spectral dependence of the dielectric function, confirming the indirect nature of the
lowest gap closely followed by a direct gap. We investigate the linear terms in \textbf{k} , some of which appear even in the absence of spin as a result of the chirality. We compare the calculated dielectric function with unpublished data by Zallen \textit{et al.}. We use the ABINIT code to calculate the frequencies of Raman and ir phonons and their dispersion relations. The densities of states of one and two phonons are also calculated. We devote the last section to present experimental data on the specific heat vs. temperature for natural samples and especially grown isotopically modified ones. These results are compared with \textit{ab initio} calculations. Generally, good agreement between experiment and \textit{ab initio} results is observed.

\begin{acknowledgments}
A.H.R. has been supported by CONACyT Mexico
under project J-59853-F and by PROALMEX/DAAD. Further computer resources
have been provided by CNS IPICYT Mexico.
A. M. acknowledges the financial support from the Spanish MCYT under grants MAT2007-65990-C03-03,  CSD2007-00045 and the supercomputer  resources provides by the Red Espa\~{n}ola de Supercomputaci$\acute{\rm{o}}$n.
We thank M. Giantomassi for the help with the calculation of optical constants and K. Syassen for help with the Kramers-Kronig transformation.
We are also indebted to N.N.  for a critical reading of the manuscript.

\end{acknowledgments}

\end{document}